\documentclass[10pt, conference, letterpaper]{IEEEtran}
\usepackage{graphicx}
\usepackage{amsfonts}
\usepackage{amsmath}
\usepackage{amssymb}
\usepackage{color}
\usepackage{graphicx,color,overpic}
\usepackage{psfrag}
\usepackage{amssymb}
\usepackage{times}
\usepackage{latexsym}
\usepackage{bm}
\usepackage{cases}
\usepackage{array}
\usepackage{fancyhdr}
\usepackage{setspace}
\usepackage{subfigure}
\usepackage{url}
\usepackage{multirow}
\usepackage{epstopdf}

\usepackage{epsfig}
\usepackage{fancybox}
\usepackage{textcomp}

\ifCLASSINFOpdf
\else
\fi
\begin{document}
\IEEEoverridecommandlockouts
\title{Selective AP-sequence Based Indoor Localization without Site Survey}

%\author{Ran Liu$^{\ddag}$, Chau Yuen$^{\ddag}$, Jun Zhao$^{\dag}$, Jindong Guo$^{\dag}$, Ronghong Mo$^{\ddag}$, Vishesh N Pamadi$^{\S}$, Xiang Liu$^{\dag}$, \\
%$^{\ddag}$ Singapore University of Technology and Design, Email: \{ran\_liu, yuenchau, ronghong\_mo \}@sutd.edu.sg \\
%$^{\dag}$ Peking University, China, Email: zhaojunpku2013@163.com, guo\_jd@pku.edu.cn, xliu@ss.pku.edu.cn \\
%$^{\S}$ Vellore Institute of Technology (VIT University), Vellore, Tamil Nadu, India, Email: visheshnp@gmail.com
%\thanks{This work is supported by Temasek Lab and NSFC 61550110244.}}

\author{Ran Liu, Chau Yuen, Jun Zhao, Jindong Guo, Ronghong Mo, Vishesh N Pamadi, Xiang Liu
\thanks{This work is supported by Temasek Lab and NSFC 61550110244. \newline
R. Liu, C. Yuen, and R. Mo are with the Singapore University of Technology and Design, 8 Somapah Rd, Singapore 487372
{\{\tt\small ran\_liu, yuenchau, ronghng\_mo@sutd.edu.sg\}}; J. Zhao, J. Guo, and X. Liu are with the Peking University, China
{\{\tt\small  zhaojunpku2013@163.com, guo\_jd@pku.edu.cn, xliu@ss.pku.edu.cn\}};
V. N. Pamadi is with the Vellore Institute of Technology (VIT University), Vellore, Tamil Nadu, India,
{\{\tt\small  visheshnp@gmail.com\}}
}}


%Telephone: (800) 555--1212, Fax: (888) 555--1212}
\maketitle

\begin{abstract}
%\boldmath
In this paper, we propose an indoor localization system employing ordered sequence of access points (APs) based on received signal strength (RSS). 
Unlike existing indoor localization systems, our approach does not require any time-consuming and laborious site survey phase to characterize the radio signals in the environment. 
To be precise, we construct the fingerprint map by cutting the layouts of the interested area into regions with only the knowledge of positions of APs. 
This can be done offline within a second and has a potential for practical use. 
%Adopting selective ordered sequence of APs as location signature, the proposed system is more robust to signal fluctuations in indoor environments. 
The localization is then achieved by matching the ordered AP-sequence to the ones in the fingerprint map.
Different from traditional fingerprinting that employing all APs information, 
we use only selected APs to perform localization, due to the fact that, without site survey, 
the possibility in obtaining the correct AP sequence is lower if it involves more APs.
%using all APs available in the Wifi network, we further select some APs out of the total APs to perform localization by employing K-means clustering algorithm, due to the fact that more APs involved in localization, 
%Rather than using all APs available in the Wifi network, we further select some APs out of the total APs to perform localization by employing K-means clustering algorithm, due to the fact that more APs involved in localization, 
%lower possibility in obtaining a correct AP sequence. 
Experimental results show that, the proposed system achieves localization accuracy $<5m$ with an accumulative density function (CDF) of $50\%$ to $60\%$ depending on the density of APs. 
Furthermore, we observe that, using all APs for localization might not achieve the best localization accuracy, e.g. in our case, $4$ APs out of total $7$ APs achieves the best performance. 
In practice, the number of APs used to perform localization should be a design parameter based on the placement of APs.
\end{abstract}

%\begin{IEEEkeywords}
%indoor localization, RSSI, Wifi, ordering
%\end{IEEEkeywords}

% \fi
\IEEEpeerreviewmaketitle

%%%%%%%%%%%%%%%%%%%%%%%%%%%%%%%%%%%%%%%%%%%%%%%%%%%%%%%%%%%%%%%%%%%%%%%%%%%%%%%%%%%%%%%%%%%%%%%%%%%%%%%%
\section{Introduction}
\label{section:introduction}
%%%%%%%%%%%%%%%%%%%%%%%%%%%%%%%%%%%%%%%%%%%%%%%%%%%%%%%%%%%%%%%%%%%%%%%%%%%%%%%%%%%%%%%%%%%%%%%%%%%%%%%%
%Over the decades, GPS has been the dominant technology to provide location information with satisfactory accuracy for content-aware applications~\cite{Ref1}\cite{Ref2}. 
%However, GPS can only function well in outdoor environments with direct visibility to the satellites, e.g. line of sight (LOS) transmission. 
%When indoor environments are concerned, weak GPS signals inside buildings as well as dense multipath problems dramatically degrade the performance of GPS~\cite{Ref3}\cite{Ref4}.

%On the other hand, there is an increasing interest in indoor localization due to the rapid growth of location-aware applications requesting room-level localization services.
Over the decades, there is an increasing interest in indoor localization due to the rapid growth of location-aware applications requesting room-level localization services. 
%Furthermore, with ubiquitous availability of WLAN infrastructures in indoor environments, 
%Wifi-based indoor localization technologies have received intensive attention due to its cost effectiveness, easy deployment, and simple hardware implementation~\cite{Ref5}\cite{Ref6}. 
Wifi-based localization systems have been shown to be successful in different scenarios, 
with a state-of-art Wifi-based localization system being reported in the literature to achieve localization accuracy below 10\,m \,\cite{Recent_advance_wireless}.
%Different localization systems might use different measurements of radio signals to locate the objects of interest, 
%including time-of-arrival (TOA), time-difference-of-arrival (TDOA), angle-of-arrival (AOA), and received signal strength (RSS)~\cite{Ref10}. 
%A majority of indoor localization systems adopt RSS as a metric since it is readily to be extracted in many off-the-shelf devices such as Wifi, RFID, and Bluetooth.
The indoor localization systems can be classified into two categories, the model-based approach\,\cite{Model_based_appraoch} and the fingerprinting-based approach\,\cite{fingerpringting_approach} \cite{Centaur}. 
The model-based approach has to rely on a model to characterize the propagation of radio signals. 
%In practice, due to severe multipath and numerous reflecting surfaces~\cite{Ref11} in indoor environments, it is very different to find such as model. 
%Therefore, the accuracy of triangulation-based approaches is much worse than the fingerprinting-based approaches. 
In contrast, fingerprinting-based approach utilizing site survey of RSS in the area of interest achieves better localization accuracy, 
since it uses a database in which the impact of environments has been underlined in the construction of the database. 
%The key idea of fingerprinting-based approaches is to partition the interested area into small regions, with each uniquely determined by a location signature and a reference position. 
%Typically, the fingerprinting-based approach consists of two phases, a surveying phase and a fingerprint matching phase. 
%The first phase involves an offline site survey process, in which engineers collect the RSS from multiple transmitters for every region of the concerned area to form a location signature, and build a fingerprint map accordingly. 
%In the second phase, the device is located to a reference position by matching its measured location signature against the signatures in the fingerprint map. 
%The location signature might have different representations, e.g. a set of absolute RSS values or stochastic description of RSS values~\cite{Ref12}, while the matching algorithm might differ in various systems.

Unfortunately, there are some drawbacks in the traditional fingerprinting-based approaches, e.g. time consuming and labor intensive site survey to construct and update the fingerprint map \cite{Mo_loc} \cite{Crowdsourcing}. To overcome this problem, 
we propose an approach to avoid the laborious phase to collect measurements in environments. 
In particular, we construct the fingerprint map by cutting the layouts of the interested area into regions with the knowledge of positions of access points.
The resulting fingerprint map is in essence a set of small connecting regions, obtained by cutting the lines connecting APs in the interested area in the middle without laborious site survey. 
This is equivalent to assuming a radio propagation model universal to all environments. 
The resulting fingerprint map can be easily constructed, with only the knowledge of locations of APs but regardless of the layout of interested area. 
Each region is then associated to an unique ordered sequence of APs (location signature). 
As a consequence, the proposed system can be deployed rapidly, while ensuring satisfactory localization accuracy competitive to existing systems. 
The proposed system might be applied to scenarios where the on-site data collection is not possible, for example disaster areas.

With this fingerprint map, the location signature is represented by an ordered sequence of APs according to RSS. 
In this case, we utilize the relative RSS among APs rather than the absolute RSS of each AP. 
%Since our approach does not need any laborious phase to collect the RF measurements beforehand, 
%thus can be applied to scenarios where new transmitters are installed without recollecting the measurements, which is required for the traditional fingerprinting-based approaches.
Note that, although we classify our new system as fingerprinting-based approach, unlike existing fingerprinting-based approaches, 
it does not require any time-consuming site survey phase.
%, which will greatly improve the localization accuracy. 
The overwhelming overhead in constructing and updating the fingerprint map are also considerably reduced due to the less number of reference positions needed to characterize the fingerprint map.

However, when AP-sequence is used as location signature, with increasing number of APs involved in localization, 
the probability to have a correct measured signature decreases (since no site survey is performed), and 
resulting in matching against a wrong signature in the map and thus a wrong reference position. 
We therefore argue that the AP-sequence localization system based on employing RSS from all present transmitters might not be efficient. 
Motivated by this observation, in our proposed system, the location signature is represented by selected APs rather than all the available APs in the range. 
The rationale behind is that with less number of RSS, higher probability to obtain a correct measured signature is expected. 
Specifically, we propose a K-means~\cite{Kmeans} based algorithm to efficiently select APs out of the total available APs for localization.

The rest of the paper is organized as follows. 
The system model underlined is presented in Sect.~\ref{section:system_model}. 
The proposed method to perform location estimation is illustrated in Sect.~\ref{section:localization_method}. 
Experiments to evaluate the performance of the proposed system is carried out in Sect.~\ref{section:experimental_results}. 
Last, we conclude this paper in Sect.~\ref{section:conclusion}.

\begin{figure}
\centering
\includegraphics[width=0.4\textwidth]{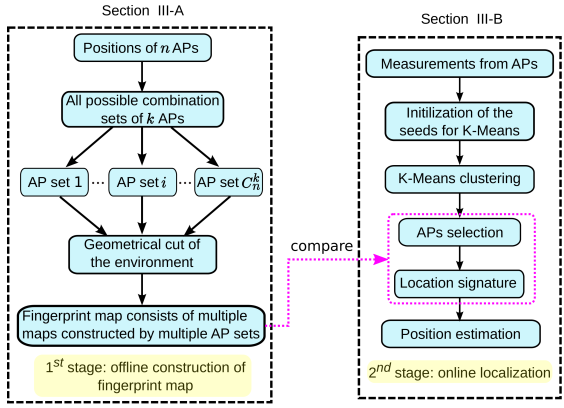}
\caption{An overview of proposed selective AP-sequence indoor localization system.}
\label{fig:system_overview}
\vspace{-0.05in}
\end{figure}

%%%%%%%%%%%%%%%%%%%%%%%%%%%%%%%%%%%%%%%%%%%%%%%%%%%%%%%%%%%%%%%%%%%%%%%%%%%%%%%%%%%%%%%%%%%%%%%%%%%%%%%%
\section{System Model}
\label{section:system_model}
%%%%%%%%%%%%%%%%%%%%%%%%%%%%%%%%%%%%%%%%%%%%%%%%%%%%%%%%%%%%%%%%%%%%%%%%%%%%%%%%%%%%%%%%%%%%%%%%%%%%%%%
Assume an indoor environment with windows, partitions, furniture, and equipments. 
%Dense multipath is expected to be present during the propagation of radio signals from APs to users. 
%An IEEE 801.11 wireless network with multiple APs is often deployed within the area. 
Users equipped with Wifi receivers are able to scan APs in the proximity and measure RSS.
%The APs are assumed to transmit radio signals at identical power level in the long run but %instantaneous power adaptation is allowed, since in advanced Wifi network, dynamic power %control mechanism has been adopted thus the transmission power of APs can be dynamically %changed based on the user density and the traffic loads. This instantaneously power %adjustment is impossible to be reflected in fingerprinting-based localization system, since %it requires frequently updating the fingerprint map aligned to the unpredictable changes. 
%However, our system is able to address this problem, since we do not need any offline survey %phase to collect the measurements. 
%The impact of dynamic power adjustment can only be investigated by experiments.
The locations of APs are assumed to be known as a prior. 
This is reasonable since in many commercial and industrial buildings, the AP locations are predetermined to optimize the network coverage and link quality. 
In case where the AP locations are not known, there are many reported ways to effectively locate the APs in practice~\cite{Ref14}. 
%For example, Han et.al. \cite{Ref15} proposed to localize APs using the distribution of the local signal strength.

\begin{figure}
  \centering
    \subfigure[The RSS of all APs]{
\label{fig:all_ap_rss}
    \includegraphics[width=0.4\textwidth]{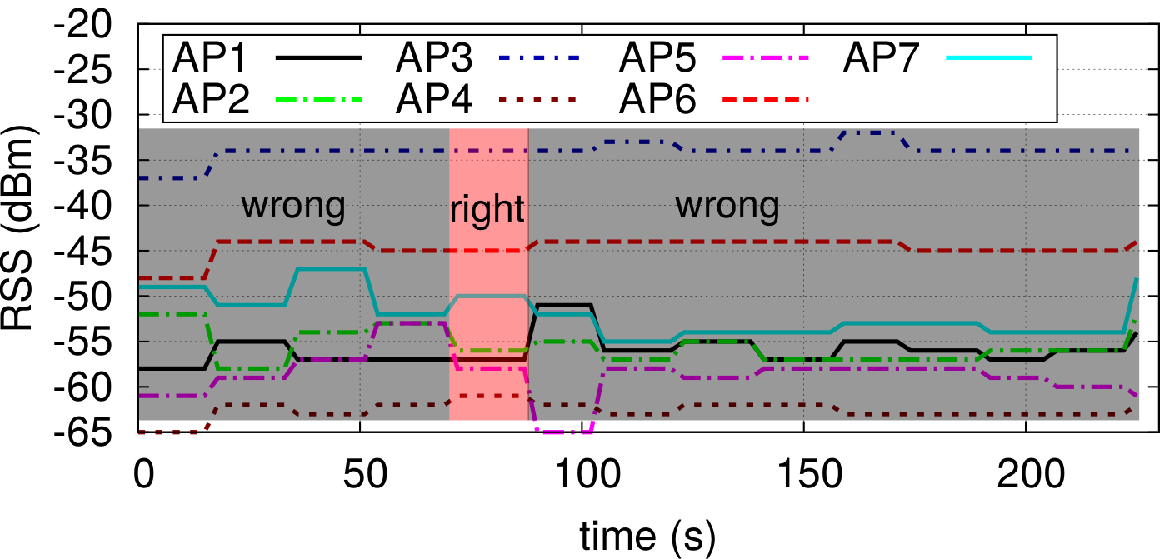}}
  \subfigure[The RSS of four APs]{
\label{fig:four_ap_rss}
        \includegraphics[width=0.4\textwidth]{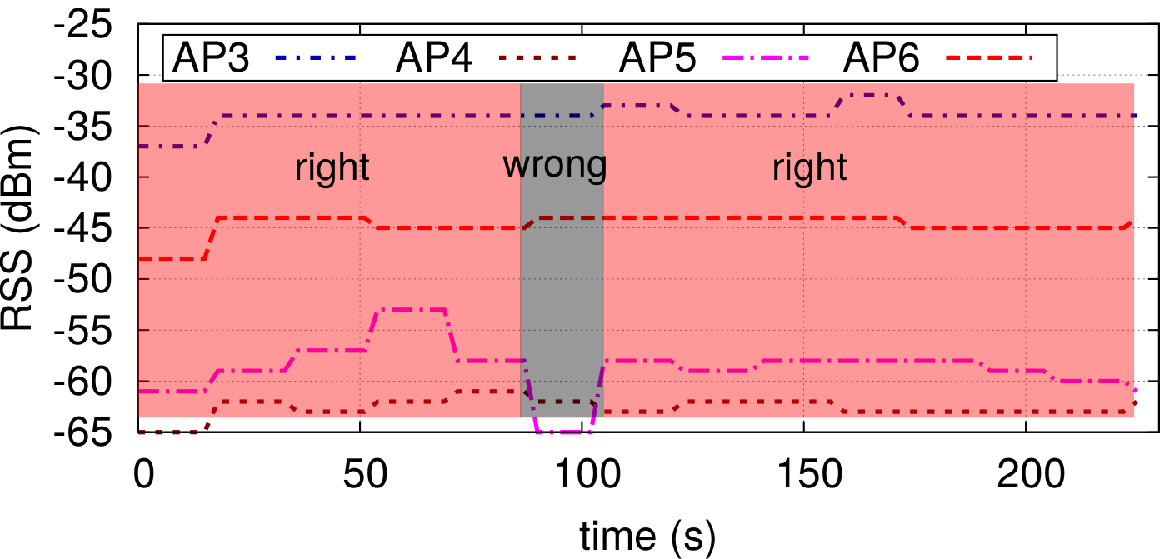}}
   \caption[RSS of the APS]
{The signal strength distribution during four minutes of all APs and four APs at one test point.}
\label{fig:RSS}
\vspace{-0.2in}
\end{figure}

Our fingerprinting-based localization system essentially consists of two phases. 
First, 
%fingerprint map which concomprised of small regions determined by the location signatures is constructed. 
we obtain the offline fingerprint map consisting of multiple maps, 
each of which is constructed by using a combination set of APs used for localization. 
In total, there will be $C_n^k$ fingerprint maps, 
assuming we have $n$ APs in the environments, and $k$ APs are chosen for localization. 
%The reason why we want to use $k$ APs for localization is explained for the next section.
Second, the user measures RSS from APs in the range and forms the measured location signature, 
then queries the corresponding fingerprint map to retrieve its location information by employing a matching method. 

During the first phase (see Fig. \ref{fig:system_overview}), 
%we construct the offline fingerprint map using all combination sets of APs choosen for localization.
we construct the fingerprint map by partitioning the environment into a set of regions, with each associated with a location signature. 
The traditional fingerprinting-based approach requires a laborious site survey phase to collect the radio signals in environments. 
In contrast, in this paper, by assuming a universal propagation model of AP, 
a region associated with a location signature is equivalent to a sub-area enclosed by cutting the lines connecting the APs in the middle, given the positions of APs.

During the second phase, i.e. online localization, the location signature is represented by an ordered AP sequence based on RSS. 
As an example, assume the number of available APs is $M$, the RSS of the $m^{th}$ AP is denoted as $S_m$. 
Arranging the RSS of all the $M$ APs in descending order, for example $(S_1, S_3, S_M, \cdots, S_5)$, the corresponding location signature will be given as $(1, 3, M, \cdots, 5)$.
%Since one region is represented by a reference position, the reference position needs to be determined so as to well characterize the region. Here the reference point is obtained as follows. Assume a region $k$ comprised of $N$ locations. The $n^{th}$ location $n$ of region $k$ is denoted by $(x_{k,n}, y_{k,n})$ where $x_{k,n}, y_{k,n}$ are the coordinates, the reference position $(x_k, y_k)$ for this region is given by

Note that, although there might be many APs available in the Wifi network, only a subset of available APs are used for localization. 
This is because more APs involved in the localization leads to a lower probability to obtain a matched fingerprint, as shown in Fig. \ref{fig:RSS}. 
This figure shows the variations of the signal strength from APs during a certain time at one test point. 
The correct AP sequence at this test point should be $3672154$. 
In fact, due to many influencing factors from the environment, the measured AP sequence is often wrong, as can be seen from Fig. \ref{fig:all_ap_rss}.
Furthermore, we can observe that the probability to get a right fingerprint using four APs is much higher than using all APs (i.e. seven APs).
This is why we propose to use selective-AP sequence to improve the probability to obtain a right signature. 
The subset of APs used is dynamically determined based on the measured RSS, and varies from position to position. 
An algorithm based on the $K$-means clustering will be employed for efficient and tractable selection of APs \cite{Kmeans}, 
with details shown in Section \ref{subsection:online_localization}. 
And the selected APs sequence will then match with the fingerprint map generated using the same set of APs. 

\begin{figure*}
  \centering
    \subfigure[Fingerprint map if AP1, AP3, AP5, AP7 are selected]{
\label{fig:select_1357_environment_partition}
    \includegraphics[width=0.23\textwidth]{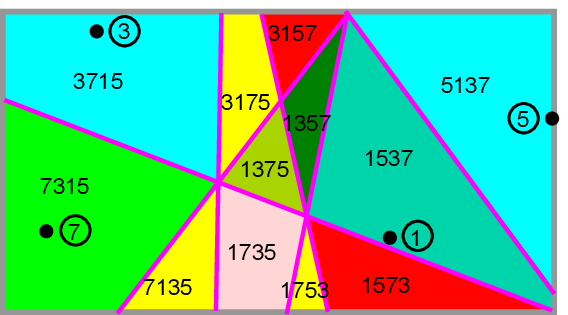}
    }
  \subfigure[Fingerprint map if AP1, AP2, AP4, AP7 are selected]{
\label{fig:select_1247_environment_partition}
        \includegraphics[width=0.23\textwidth]{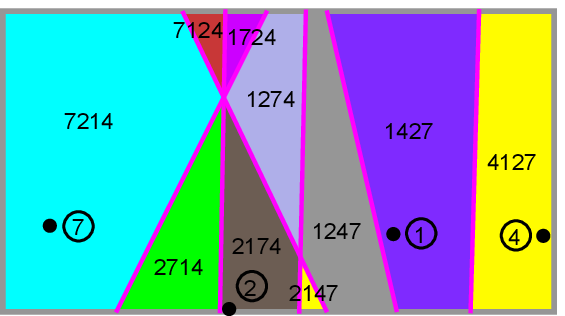}
        }
    \subfigure[Fingerprint map if AP1, AP4, AP7 are selected]{
\label{fig:select_147_environment_partition}
        \includegraphics[width=0.23\textwidth]{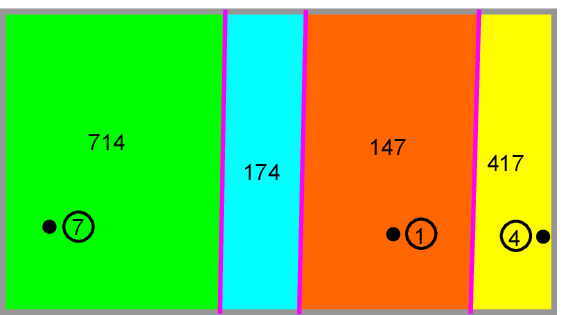}
        }
          \subfigure[Fingerprint map if AP2, AP6, AP7 are selected]{
\label{fig:select_267_environment_partition}
        \includegraphics[width=0.23\textwidth]{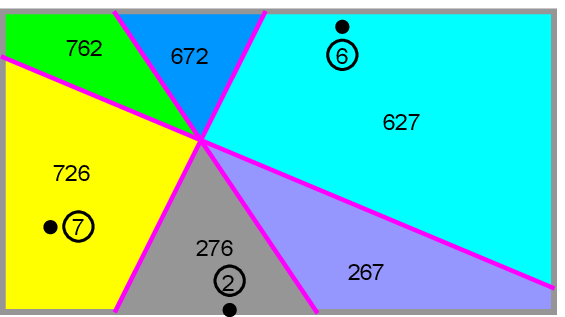}
        }
   \caption[AP fingerprint model.]
{Illustration of fingerprint maps for different combination sets of APs.}
\label{fig:AP_fingerpringting_model}
\vspace{-0.2in}
\end{figure*}

%%%%%%%%%%%%%%%%%%%%%%%%%%%%%%%%%%%%%%%%%%%%%%%%%%%%%%%%%%%%%%%%%%%%%%%%%%%%%%%%%%%%%%%%%%%%%%%%%%%%%%%%
\section{Localization Method}
\label{section:localization_method}
%%%%%%%%%%%%%%%%%%%%%%%%%%%%%%%%%%%%%%%%%%%%%%%%%%%%%%%%%%%%%%%%%%%%%%%%%%%%%%%%%%%%%%%%%%%%%%%%%%%%%%%%

The localization system proposed in this paper uses the relative RSS rather than the absolute RSS for localization and has two stages, 
offline construction of fingerprint map and online localization, as shown in Fig.~\ref{fig:system_overview}. 
Particularly, the offline stage neither needs any laborious site survey phase to collect the measurements nor requires any mathematical or analytical model to characterize the propagation of the radio signals.
The resulting fingerprint map is comprised of multiple fingerprint maps, which are constructed by various combination sets of APs.
In the online stage, a subset of APs is selected based on K-means clustering and a location signature based on RSS measurements is obtained. 
The measured location signature is then matched against the location signatures 
in the corresponding fingerprint map (e.g. if AP 124 is selected, then the AP sequence will match to the fingerprint map generated by AP 124) 
and the position information is obtained.

\subsection{Offline Construction of Fingerprint Map}
The fingerprint map is constructed by partitioning the interested area into a set of regions, with each associated with a location signature. 
Conventionally, the construction of fingerprint map requires offline site survey, involving measuring radio signals in every position in the environment. 
In contrast, in this paper, by adopting the ordered AP sequence based on RSS and assuming a universal propagation model for all environments, 
a region associated with a location signature is equivalent to a sub-area enclosed by cutting the lines connecting the APs in the middle, given the locations of APs but without any site survey. 
The shape of each region depends on the displacements of the selected APs for localization.

It is expected that, the number of regions obtained depends on the number and the placement of APs. 
Typical fingerprint maps with 4 and 3 APs are given in Fig.~\ref{fig:AP_fingerpringting_model}, 
where the locations of the APs are shown by \textcircled{n} for the $n^{th}$ AP, 
while the location signatures of different regions are denoted by ordered sets of APs, for example, $n_1 n_2 n_3 n_4$ and $n_1 n_2 n_3$, 
with $n_1, n_2, n_3, n_4$ being the APs whose RSS arranged in descending order, e.g. $RSS_{n_1}>RSS_{n_2}>RSS_{n_3}>RSS_{n_4}$. 
It can be seen that, with more APs involved, more regions are obtained, for example, $12$ regions in Fig.~\ref{fig:select_1357_environment_partition} and $6$ regions for Fig.~\ref{fig:select_267_environment_partition}.

Although we only show the maps of several sets of selected APs as examples, 
given the number of available APs in the network to perform localization, there will be multiple fingerprint maps for all possible combination sets of APs. 
Note that in this stage, the fingerprint maps for all possible combinations of APs are constructed. 
In the localization stage, the APs used to perform localization are selected dynamically using the K-means algorithm.

Given the fingerprint map, the localization accuracy of each region is computed as the average distance between all positions and the center of the region. 
%The theoretically achieved accuracy can then be derived by averaging the region localization accuracy over all regions in the map. 
%The accuracies for various combination sets of APs are shown in Fig.~\ref{fig:AP_fingerpringting_model_error}, 
%with accuracies in green color denoting the worst case accuracy, 
%while the accuracies in black color denoting the average accuracy in that region. 
Theoretically, higher localization accuracy can be achieved with more APs involved in localization. 
%For example, the average localization accuracy is $1.3m$ using all APs, while it is $2.2m$ for the set of APs $1357$ and $4.1m$ for the set of APs $147$, 
%as shown in Fig.~\ref{fig:AP_fingerpringting_model_error}. 

In practice, the measured location signature might not be correct, 
for example, as shown in Fig. \ref{fig:all_ap_rss}, the correct location signature is $(3,6,7,2,1,5,4)$ while the measured location signature is $(3,6,1,7,2,4,5)$ at $t=100s$. 
So if we use all APs for localization, the theoretical accuracy we can achieve is the best, but with high chance to make mistake. 
But if we use only 4 APs, then the accuracy is lower, but with higher chance to make it correct. 
%Therefore, in practice, AP-sequence localization with more APs is not necessary to offer better accuracy.
Therefore, in practice, AP-sequence localization with less selected APs may offer better accuracy in the absence of site survey than using all AP for sequencing. 

%%%%%%%%%%%%%%%%%%%%%%%%%%%%%%%%%%%%%%%%%%%%%%%%%%%%%%%%%%%%%%%%%%%%%%%%%%%%%%%%%%%%%%%%%%%%%%%%%%%%%%%%
\subsection{Online Localization}
\label{subsection:online_localization}
\begin{figure}
\centering
\includegraphics[width=0.3\textwidth]{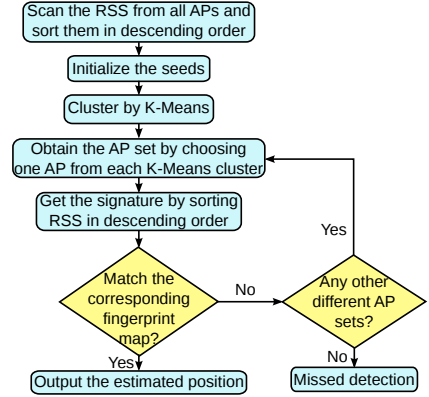}
\caption{Procedure of online localization.}
\label{fig:algorithm_detailed}
\vspace{-0.2in}
\end{figure}

%%%%%%%%%%%%%%%%%%%%%%%%%%%%%%%%%%%%%%%%%%%%%%%%%%%%%%%%%%%%%%%%%%%%%%%%%%%%%%%%%%%%%%%%%%%%%%%%%%%%%%%%
The online localization is performed in three steps, 
1) clustering APs based on their RSS by employing $K$-means clustering algorithm, 
2) selecting one AP from each K-means cluster and forming the selected APs for localization, 
and 3) position estimation. The Step 1 and Step 2 are shown in Fig.~\ref{fig:algorithm_detailed}.

\subsubsection{Clustering by K-means}
The K-means algorithm is a qualitative way to partition a group of data into a certain number of clusters (K clusters). 
In general, given a set of data $\mathbf{x}$ with $Q$ data points $\mathbf{x}=\{x_q, q=1, \cdots, Q\}$ to be partitioned into $K$ clusters, 
the goal of K-means algorithm is to assign a cluster to each data point, so that the distance from the data points to the cluster is minimized. 
Mathematically, the K-means clustering method solves the following problem:
\begin{equation}\label{eq:loc_1}
\mathrm{argmin} \sum_{k=1}^K \sum_{x_q \subseteq \mathbf{x}_k} d(x_q, \mathbf{x}_k)
\end{equation}
where $\mathbf{x}_k$ is the set of points that belong to cluster $k$ and $d(x_q, \mathbf{x}_k)$ denotes the distance between the data point $x_q$ and the $k^{th}$ cluster. 
The distance $d()$ can have various representations, for example the Euclidean distance given by $|x_q-\mu_k|$ with $\mu_k$ being the average of the data points in the $k^{th}$ cluster. 
Note that the problem (\ref{eq:loc_1}) is not trivial (in fact it is NP-hard), so the K-means algorithm only hopes to find the global minimum, probably getting stuck in a different solution.
%The K-means algorithm is indeed equivalent to defining $K$ centroids, one for each cluster, then associates each data point to its nearest centroid. 
%An iterative refinement technique is used in the algorithm by alternating between assigning data point to cluster and updating centroids of clusters, until there is no changes in data point assignment.

The $K$-means algorithm needs to be initialized. 
There are some commonly used initialization methods. 
In this paper, we simply form the initial $K$ clusters by choosing the $K$ APs with the $K$ largest RSS. 
The effect of cluster initialization will be investigated in experiments.

\subsubsection{Selection of AP set}
\begin{figure}
\centering
\includegraphics[width=0.32\textwidth]{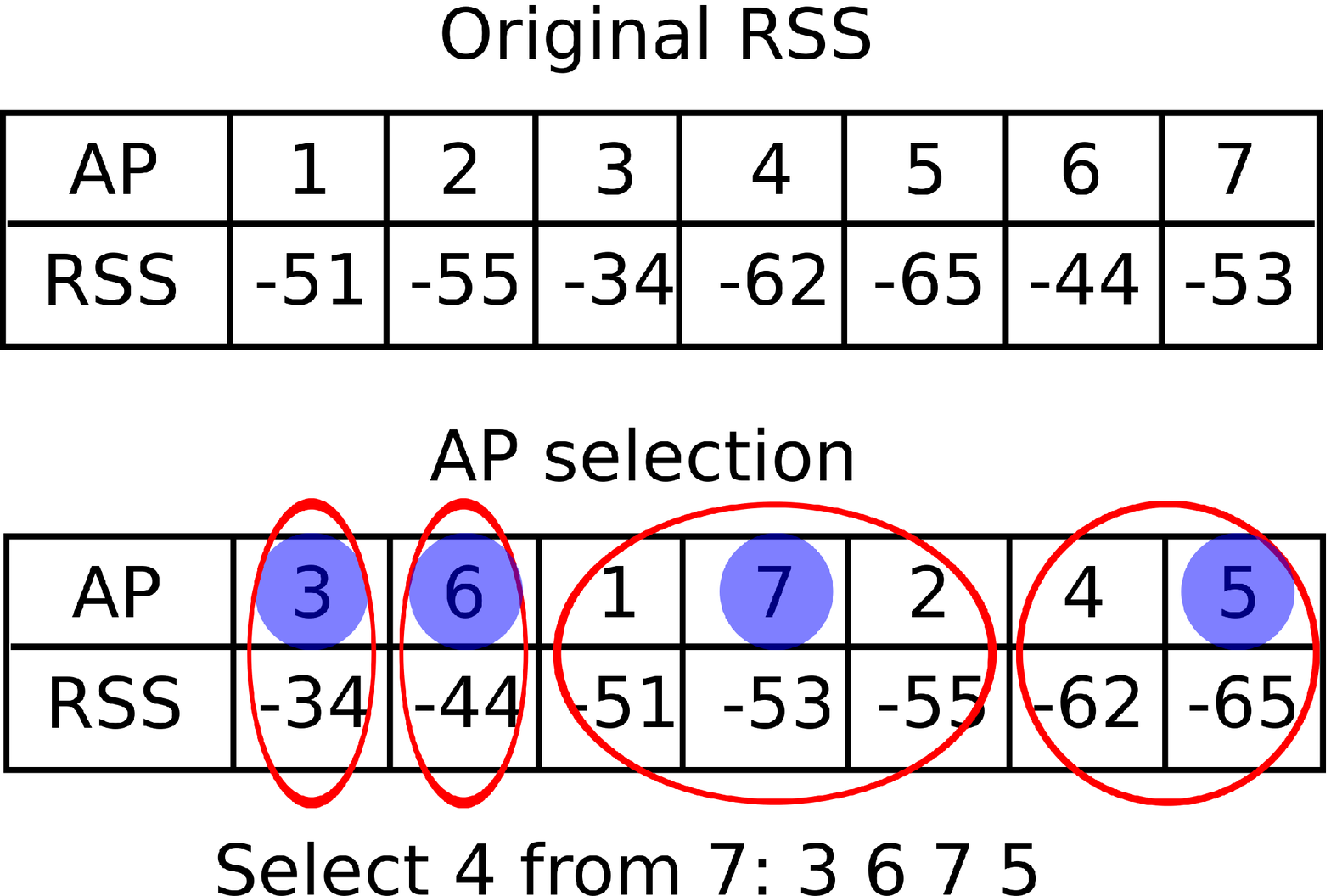}
\caption{K-means clustering and one example of the resulted AP-sequences using 4 APs (totally 7 APs in the environment).}
\label{fig:select_four}
\vspace{-0.1in}
\end{figure}

After the K-means clustering, a number of $K$ clusters are obtained. 
Next, from each cluster one AP will be selected to form a set of $K$ APs, based on the method shown in Fig.~\ref{fig:algorithm_detailed}. 
Note that, there might be multiple combinations of selected APs, since multiple APs might be in one cluster. 
In this case, we generate the AP sets by randomly selecting one AP from each cluster. 
The procedure to select a set of APs (i.e. select 4 APs from 7) can be seen in Fig. \ref{fig:select_four}. 
The K-means clustering results is denoted with the red circle, and the resulted AP sequence is denoted with a blue circle.%, i.e. $6 2 1 5$. 

\subsubsection{Position Estimation}
Once the AP set is generated, the measured location signature can be obtained by sorting the APs with RSS in descending order. 
This signature is matched against the location signatures in the fingerprint map and the location information will be retrieved. 
Note that, there is not always a location signature in the fingerprint map to match the measured location signature. 
In this case, another AP set will be selected and a new measured location signature will be constructed. 
If all the AP sets have been selected and we fail to find a matched location signature in the fingerprint map, this test point will be declared as \emph{missed detection point}.

\vspace{-0.1in}
%%%%%%%%%%%%%%%%%%%%%%%%%%%%%%%%%%%%%%%%%%%%%%%%%%%%%%%%%%%%%%%%%%%%%%%%%%%%%%%%%%%%%%%%%%%%%%%%%%%%%%%%
\section{Experiments}
\label{section:experimental_results}
%%%%%%%%%%%%%%%%%%%%%%%%%%%%%%%%%%%%%%%%%%%%%%%%%%%%%%%%%%%%%%%%%%%%%%%%%%%%%%%%%%%%%%%%%%%%%%%%%%%%%%%%
\begin{figure}
  \centering
    \subfigure[]{
\label{fig:experimental_set_up_dover}
    \includegraphics[width=0.4\textwidth]{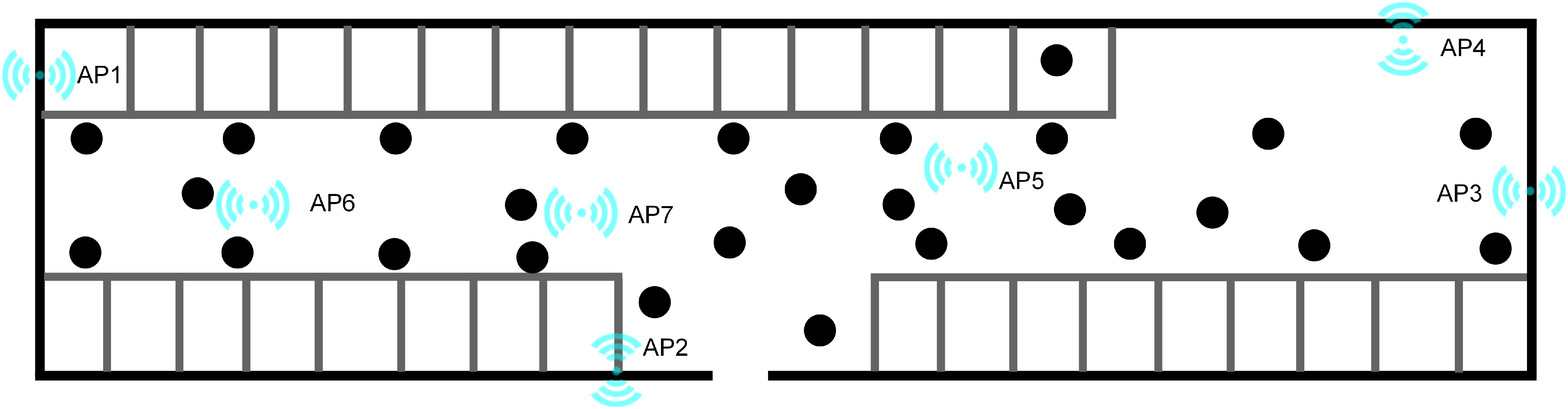}}
  \subfigure[]{
\label{fig:experimental_set_up_ECC}
        \includegraphics[width=0.3\textwidth]{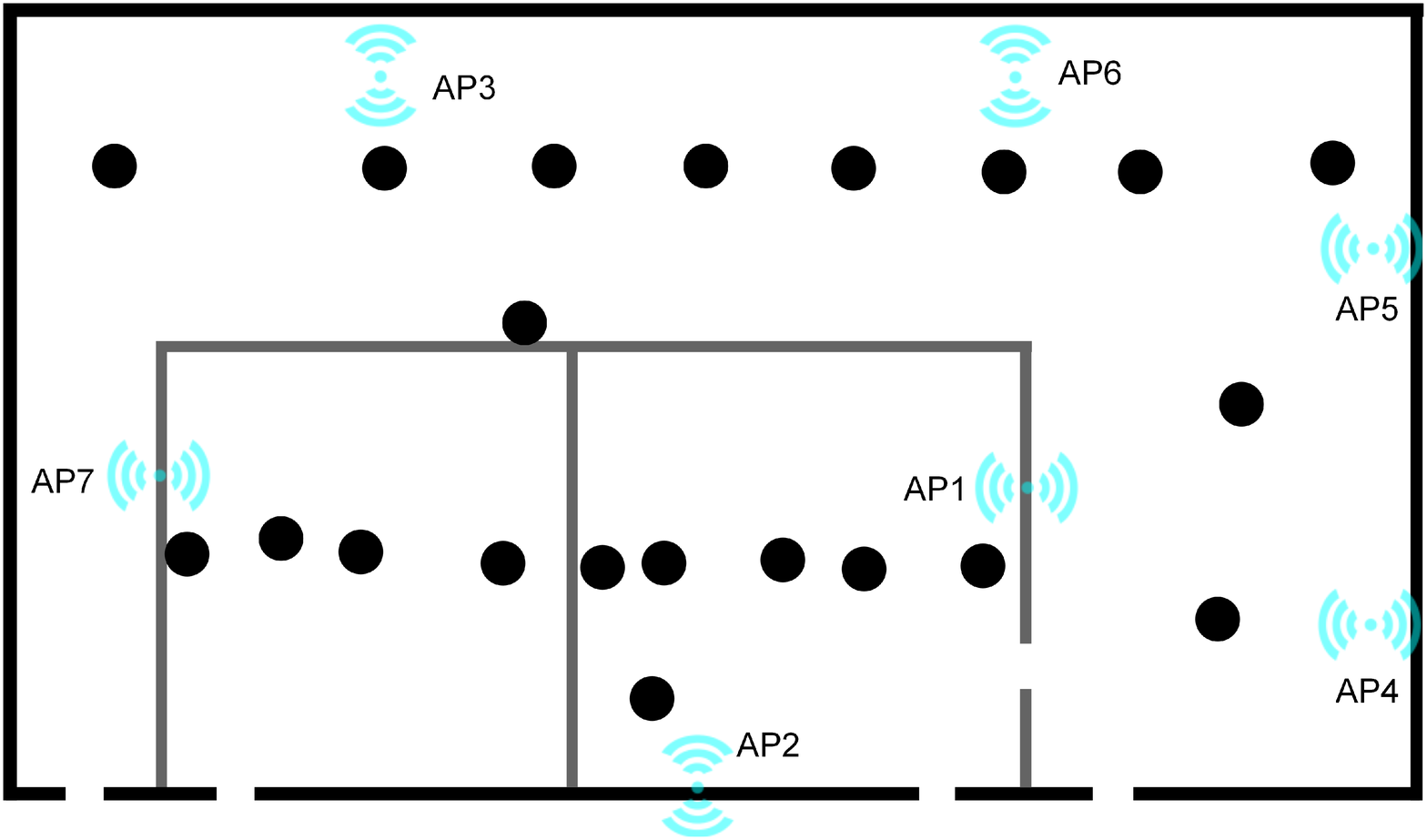}}
   \caption[Experimental setups]{Experimental setups. Black dots indicate the test points. 
(a) Experimental setup at Dover Campus with a size of (60m$\times$40m);
(b) Experimental setup at East Cost Campus (ECC) with a size of (25m$\times$14m).}
\label{fig:experimental_setups}
\vspace{-0.1in}
\end{figure}

\subsection{Experimental Setups}
We evaluate the performance of our proposed localization system by experiments, with the two setups given in Fig. \ref{fig:experimental_setups}. 
Both environments consist of concrete walls (marked with black in Fig. \ref{fig:experimental_setups}), 
soft room partitions with windows (marked with grey in Fig. \ref{fig:experimental_setups}), furniture, and equipments. 
The setup in ECC also has concrete poles inside the room. 
The setup in Dover is a room of 60m$\times$40m, while the setup in ECC is a room of 25m$\times$14m. 
The proposed system is implemented using smart phones (Sony Z2) with Android version 4.4.2.
We install 7 APs (ASUS RT-N12HP Wireless N3002) in both environments. 
The APs provide RSS values ranging from -70 dBm to -30 dBm. 
If there is no detection of the access point, we set the RSS to -100 dBm. 
At Dover campus, we collected the RSS measurements at 27 test points (see the dark dots in Fig.\ref{fig:experimental_set_up_dover}), 
while at ECC campus, which is much smaller, we recorded the measurements at 20 test points. 
For each test point, the RSS values were measured every 300~ms and the observation time is 1~min.
%It is well known that, the RSS of Wifi signals experience dense multipath and severely fluctuate in indoor environments. 
%To overcome this problem, the RSS used for localization is obtained by averaging the RSS observed over a time duration to leverage the RSS variations. %The time duration will be a design parameter based on the performance requirements.

The time used to generate the offline fingerprint map under different numbers of APs used for localization are listed in Table \ref{table:time_vs_APS}. 
The experiments were conducted using an Intel Core i5-4200M@2.50GHz CPU with 4 GB RAM.
We use a grid-based representation to compute the fingerprint map. 
For our implementation, the size of the grid is set to 0.2~m. 
%As can be seen in Table \ref{table:time_vs_APS}, constructing the fingerprint map with four APs takes more time than other combination sets of APs. 
%This is because we have to construct $C_7^4=35$ fingerprint maps in this case. 
As can be seen in Table \ref{table:time_vs_APS},
we can generate all 35 fingerprint map within one second (to be exact 26.3\,ms) for the case of 4 APs, 
which is negligible when compared to the laborious site survey phase in the traditional fingerprinting-based approaches.
%we can generate the fingerprint map within one second (for example 26.3 $ms$ for the case of 4 APs) 
%which can be ignored as compared to the laborious site survey phase in the traditional fingerprinting-based approaches.
%Moreover, the fingerprint map can be generated within one second which can be ignored as compared to the laborious site survey phase in the traditional fingerprinting-based approaches.

%Since the sensor model is generated in an off-line fashion, the computational time is not an issue in our case. 

\begin{table}
\centering
\renewcommand{\arraystretch}{1.1}
\caption[The average time (in milliseconds) consumed to construct the fingerprint map of Dover Campus.]
{Number of fingerprint maps needs to be constructed and average time (in milliseconds) consumed to construct the fingerprint map using different number of selected APs out of total 7 APs. }
\label{table:different_grid_size}
\centering
%\begin{tabular}{|c|p{0.3cm}<{\centering}|c|c|c|c|c|}
 \begin{tabular}{|p{3.4cm}<{\centering}|p{0.3cm}<{\centering}|p{0.45cm}<{\centering}|p{0.45cm}<{\centering}|p{0.45cm}<{\centering}|p{0.3cm}<{\centering}|p{0.3cm}<{\centering}|}
%\begin{tabular}{|c|p{1.5cm}<{\centering}|p{2.0cm}<{\centering}|p{2.0cm}<{\centering}|p{1.2cm}<{\centering}|p{2cm}<{\centering}|p{2.1cm}<{\centering}|}
\hline
 \multirow{1}{*}{\centering Number of selected APs}  &
  \multirow{1}{*}{\centering 2}&
 \multirow{1}{*}{\centering 3}&
 \multirow{1}{*}{\centering 4}&
 \multirow{1}{*}{\centering 5} &
 \multirow{1}{*}{\centering 6}&
 \multirow{1}{*}{\centering 7}\\
\hline
 \multirow{1}{*}{\centering Number of maps}  &
  \multirow{1}{*}{\centering 21}&
 \multirow{1}{*}{\centering 35}&
 \multirow{1}{*}{\centering 35}&
 \multirow{1}{*}{\centering 21} &
 \multirow{1}{*}{\centering 7}&
 \multirow{1}{*}{\centering 1}\\

 \multirow{1}{*}{\centering Time used [ms]}  &
  \multirow{1}{*}{\centering 9.4}&
 \multirow{1}{*}{\centering 21.8}&
 \multirow{1}{*}{\centering 26.3}&
 \multirow{1}{*}{\centering 25.4} &
 \multirow{1}{*}{\centering 9.2}&
 \multirow{1}{*}{\centering 1.3}\\
\hline
%\centering{Number of maps}&\centering{21}&\centering{35}&\centering{35}&\centering{21}&\centering{7}&\centering{ 1 }\\
%\hline
%\centering{Time used [ms]}&\centering{9.4}&\centering{21.8}&\centering{26.3}&\centering{25.4}&\centering{9.2}&\centering{1.3}\\

\end{tabular}
\label{table:time_vs_APS}
\vspace{-0.2in}
\end{table}

\subsection{Results and Analysis}
The proposed selective AP-sequence based indoor localization system is evaluated in terms of localization accuracy, 
which is defined as the distance between the estimated position and the test points in meter. 
The experimental results are shown in Fig.~\ref{fig:experimental_results}.

Fig.~\ref{fig:influence_seeds_dover} and Fig.~\ref{fig:influence_seeds_ECC} present the localization accuracy in terms of cumulative density function (CDF), 
given different initial seeds of K-means algorithm. 
The CDFs of accuracy obtained with different numbers of APs selected for localization are shown in Fig.~\ref{fig:influence_ap_selection_dover} and Fig.~\ref{fig:influence_ap_selection_ECC}, 
whereas Fig.~\ref{fig:influence_time_duration} gives the CDFs of accuracy with various observation time.

%%%%%%%%%%%%%%%%%%%%%%%%%%%%%%%%%%%%%%%%%%%%%%%%%%%%%%%%%%%%%%%%%%%%%%%%%%%%%%%%%%%%%%%%%%%%%%%%%%%%%%%%
\subsubsection{Impact of initial seeds of K-Means algorithm}
\begin{figure*}
  \centering
   \subfigure[]{
\label{fig:Snapshot of_setup at ECC}
        \includegraphics[width=0.32\textwidth]{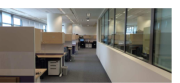}}
    \subfigure[]{
\label{fig:influence_seeds_dover}
    \includegraphics[width=0.32\textwidth]{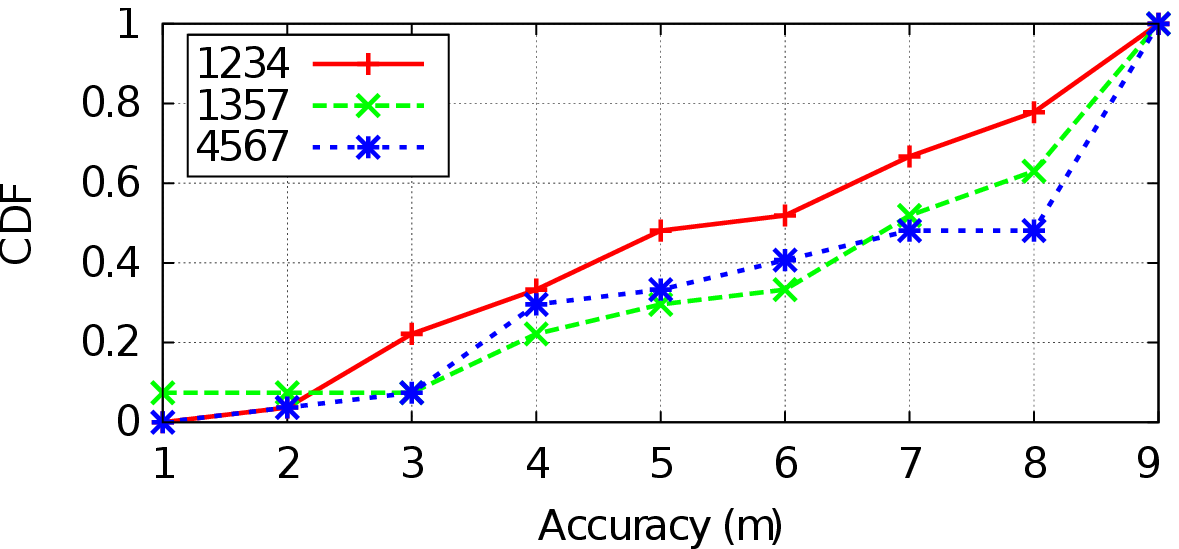}}
  \subfigure[]{
\label{fig:influence_seeds_ECC}
        \includegraphics[width=0.32\textwidth]{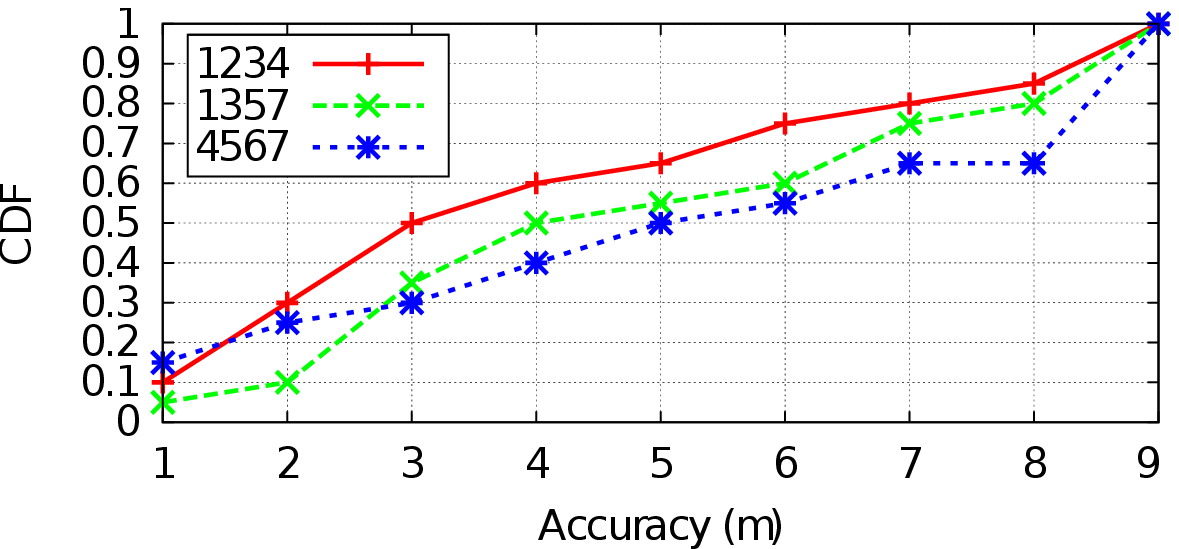}}
            \subfigure[]{
\label{fig:influence_ap_selection_dover}
    \includegraphics[width=0.32\textwidth]{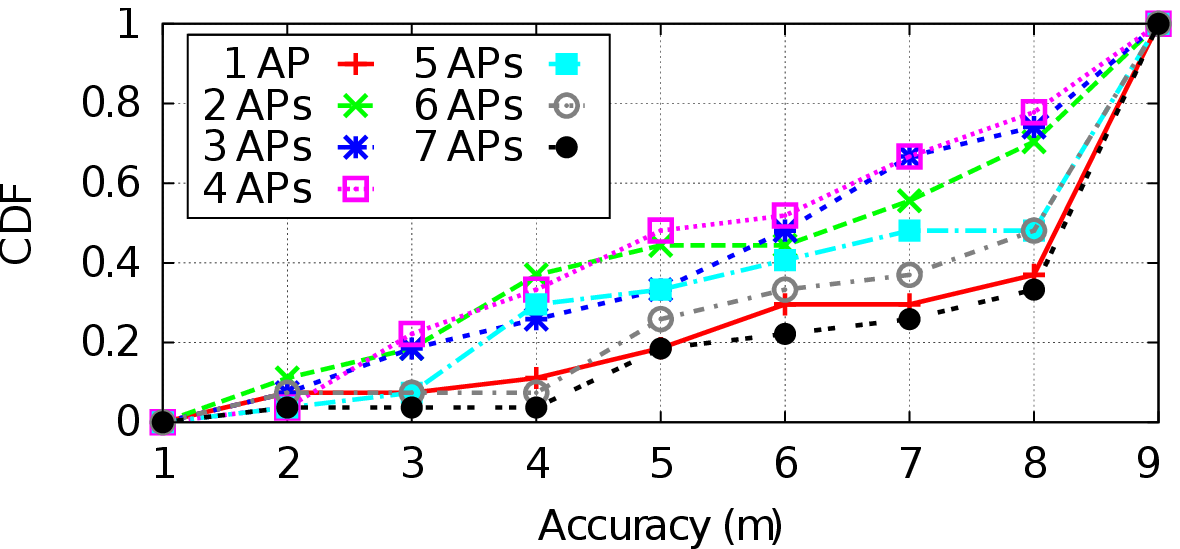}}
  \subfigure[]{
\label{fig:influence_ap_selection_ECC}
        \includegraphics[width=0.32\textwidth]{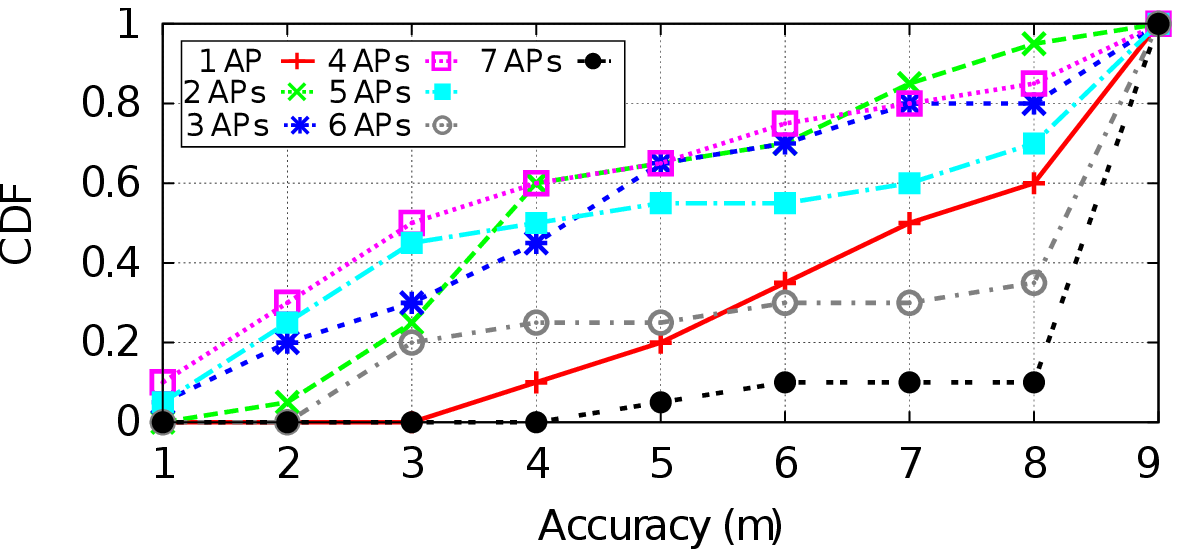}}
        \subfigure[]{
\label{fig:influence_time_duration}
        \includegraphics[width=0.32\textwidth]{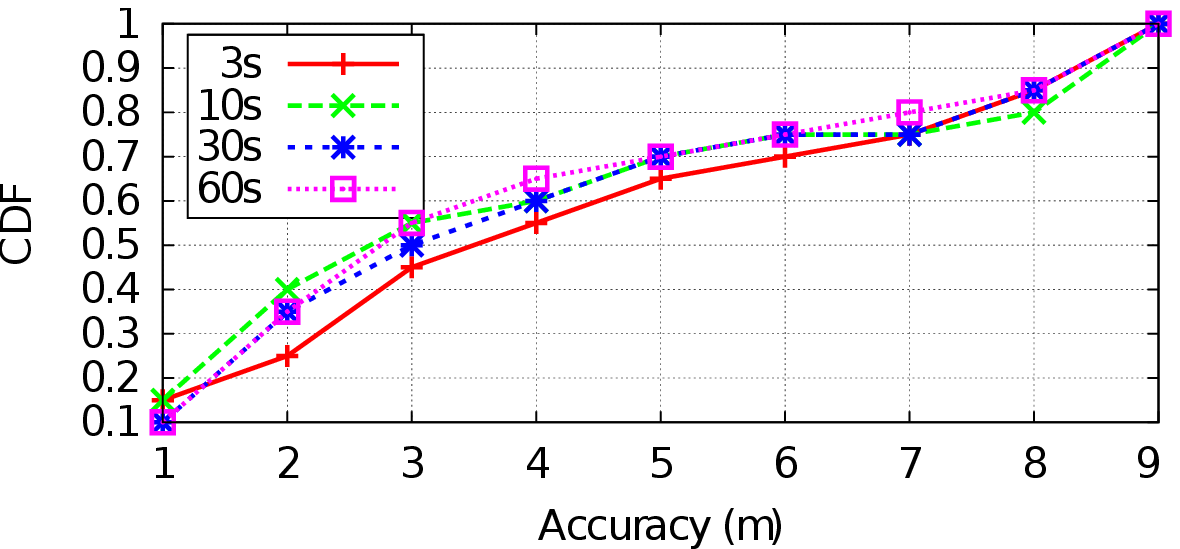}}               
   \caption[Mean Localization Accuracy under the Impact of the numbers of the slected APs.]
{(a): Snapshot of setup at ECC;
(b)\,(c): localization accuracy with various initial seeds at Dover and ECC respectively;
(d)\,(e): localization accuracy with various numbers of APs selected for localization at Dover and ECC respectively; 
(f): localization accuracy for various time durations at ECC.
}
\label{fig:experimental_results}
\vspace{-0.15in}
\end{figure*}

%%%%%%%%%%%%%%%%%%%%%%%%%%%%%%%%%%%%%%%%%%%%%%%%%%%%%%%%%%%%%%%%%%%%%%%%%%%%%%%%%%%%%%%%%%%%%%%%%%%%%%%%
In the K-means clustering, the choice of APs in the initial K clusters affects the generation of AP sets and thus the localization accuracy. 
Given $4$ APs out of $7$ APs used for localization, the accuracy with different initial seeds is shown in Fig.~\ref{fig:influence_seeds_dover} and Fig.~\ref{fig:influence_seeds_ECC} for the two experimental setups, 
where the APs with RSS ordered in $n_1n_2n_3n_4$ are selected. 
It can be seen that, given the current placement of the $7$ APs, the AP set with initial seed of $1234$ has the best accuracy for both setups.
The reason might be that, stronger signal strength is much more stable and reliable to reflect the position of the interested user, 
whereas weaker signal strength brings more uncertainty in position interpretation. 
The AP set with initial clusters of stronger RSS is therefore more robust to RSS fluctuations and a correct AP-sequence can be obtained. 
A CDF of $50\%$ can be achieved for accuracy $<5m$ in both setups.

To our best knowledge, so far there is no Wifi fingerprinting indoor localization system based on RSS without offline site survey to construct the fingerprint map. 
In the near future, we would like to compare the performance of our approach with the state-of-the-art techniques.

%all pictures here
%end all pictures here

%%%%%%%%%%%%%%%%%%%%%%%%%%%%%%%%%%%%%%%%%%%%%%%%%%%%%%%%%%%%%%%%%%%%%%%%%%%%%%%%%%%%%%%%%%%%%%%%%%%%%%%%
\subsubsection{Impact of the number of APs selected for localization}
%%%%%%%%%%%%%%%%%%%%%%%%%%%%%%%%%%%%%%%%%%%%%%%%%%%%%%%%%%%%%%%%%%%%%%%%%%%%%%%%%%%%%%%%%%%%%%%%%%%%%%%%
The localization accuracy achieved with different number of APs selected is shown in Fig.~\ref{fig:influence_ap_selection_dover} and Fig.~\ref{fig:influence_ap_selection_ECC}. 
It is observed that, the accuracy obtained using $7$ APs for localization is not the best, 
although theoretically using $7$ APs gives the best performance. 
This is because that, with more APs involved, it is not always likely to have a correct AP-sequence in the fingerprint map, resulting in performance degradation.

\begin{table}
 \centering
 \caption[Successful rate]
 {Probability of missed detection (7 APs in total).}
 \centering
 \begin{tabular}{|p{2.5cm}<{\centering}|p{0.3cm}<{\centering}|p{0.3cm}<{\centering}|p{0.3cm}<{\centering}|p{0.45cm}<{\centering}|p{0.45cm}<{\centering}|p{0.45cm}<{\centering}|p{0.45cm}<{\centering}|}
 \hline
 No. APs used &1&2&3&4&5&6&7
 \\\hline
 Dover& 0&0&0&0&0.15&0.41&0.63\\
 \hline
 ECC& 0&0&0&0&0.2&0.4&0.85\\
 \hline
 \end{tabular}
 \label{table:successful_rate_ap_selection}
 \vspace{-0.2in}
 \end{table}

We also investigate the probability of \emph{missed detection} given different numbers of APs used for localization, 
as shown in Table \ref{table:successful_rate_ap_selection}. 
Here \emph{missed detection} means we could not find any matched fingerprint in the corresponding offline fingerprint maps (see Fig.\,\ref{fig:algorithm_detailed}).
It is observed that, with more APs used for localization, higher possibility of \emph{missed detection} is expected. 
In our examples, when all 7 APs are used, $63\%$ and $85\%$ \emph{missed detection} rate are observed, while when 4 out of 7 APs are used, zero \emph{missed detection} can be achieved.

%%%%%%%%%%%%%%%%%%%%%%%%%%%%%%%%%%%%%%%%%%%%%%%%%%%%%%%%%%%%%%%%%%%%%%%%%%%%%%%%%%%%%%%%%%%%%%%%%%%%%%%%
\subsubsection{Impact of time duration to record measurements}
%%%%%%%%%%%%%%%%%%%%%%%%%%%%%%%%%%%%%%%%%%%%%%%%%%%%%%%%%%%%%%%%%%%%%%%%%%%%%%%%%%%%%%%%%%%%%%%%%%%%%%%%
The localization accuracy with various time durations (e.g. $3s$, $10s$, $30s$, and $60s$) to record the RSS measurements, is shown in Figure \ref{fig:influence_time_duration}. 
As can be seen from this figure, a longer observation time gives a better localization accuracy. 
Since the Wifi signal is severely affected by surroundings, with insufficient measurements (i.e. shorter observation time), 
we are not able to get a robust estimation of the signal strength, which will lead to a worse localization accuracy. 
However, for the real-time applications, time consumption will be a key concern, 
therefore we propose to use probabilistic-based approaches (e.g. particle filter) as our future work to deal with the uncertainties.

\vspace{-0.1in}
%\subsubsection{Comparison with state-of-the-art}
%%%%%%%%%%%%%%%%%%%%%%%%%%%%%%%%%%%%%%%%%%%%%%%%%%%%%%%%%%%%%%%%%%%%%%%%%%%%%%%%%%%%%%%%%%%%%%%%%%%%%%%%
\section{Conclusions}
\label{section:conclusion}
This paper studies an indoor localization system employing selective ordered sequence of APs based on RSS without site survey. 
The proposed system is comprised of an offline phase to generate the fingerprint map and an online phase to dynamically select the AP as well as retrieve the location information from the fingerprint map. 
In contrast to existing indoor localization systems where time-consuming and laborious site survey is necessary, 
the fingerprint map of the proposed system is easily constructed by cutting the layouts of the interested area into regions with only the knowledge of positions of APs 
and assuming a universal propagation model. 
Employing ordered sequence of APs as location signature, the proposed system is more robust to signal fluctuations in indoor environments. 
Rather than using all APs available in the Wifi network, we use a selected set of APs to perform localization by employing K-means clustering algorithm. 
Experimental results show that, the proposed system achieves localization accuracy $<5m$ with a CDF of $50\%$ to $60\%$ depending on the density of APs. 
Furthermore, we provide experimental results in terms of localization accuracy and demonstrate that our proposed selective AP sequence performs better than using all APs for localization. 
The number of APs used to perform localization should be a design parameter based on the placement of APs, which will be further investigated in our future work. 
Another direction of our future work will be the comparison of our approach to the state-of-the-art.
%%%%%%%%%%%%%%%%%%%%%%%%%%%%%%%%%%%%%%%%%%%%%%%%%%%%%%%%%%%%%%%%%%%%%%%%%%%%%%%%%%%%%%%%%%%%%%%%%%%%%%%%

%%%%%%%%%%%%%%%%%%%%%%%%%%%%%%%%%%%%%%%%%%%%%%%%%%%%%%%%%%%%%%%%%%%%%%%%%%%%%%%%%%%%%%%%%%%%%%%%%%%%%%%%
%\section*{Acknowledgment}
%\label{section:acknowledgment}
%%%%%%%%%%%%%%%%%%%%%%%%%%%%%%%%%%%%%%%%%%%%%%%%%%%%%%%%%%%%%%%%%%%%%%%%%%%%%%%%%%%%%%%%%%%%%%%%%%%%%%%%

\end{document}